\documentclass[12pt]{article}
\usepackage[english]{babel}
\usepackage{amsmath, amsfonts, amssymb, amsthm, dsfont, graphicx, tabularx, adjustbox, graphics, bbm, mathrsfs, accents, upgreek, mathtools, multirow, nameref, natbib, setspace, thmtools}
\usepackage{enumitem}
\usepackage[font=footnotesize,justification=centering]{caption}
\usepackage{subcaption}

\usepackage{tikz}
\usetikzlibrary{fit,positioning,arrows,automata,calc}
\usepackage{verbatim}
\usepackage{xr}
\usepackage[normalem]{ulem}
\usepackage{subfiles}

\usepackage{tabularx,booktabs}

\newcommand{\abs}[1]{\vphantom{#1}\left\lvert\smash[t]{#1}\right\rvert}
\newcommand{\norm}[1]{\vphantom{#1}\left\lVert\smash[t]{#1}\right\rVert}
\newcommand{\bb}{\mathbf{b}}

\newcommand{\alphahat}{{\widehat{\alpha}}}
\newcommand{\bbeta}{{\boldsymbol{\beta}}}
\newcommand{\bbetao}{{\boldsymbol{\beta}^o}}
\newcommand{\bbetahat}{{\widehat{\boldsymbol{\beta}}}}
\newcommand{\bbetacheck}{{\widecheck{\boldsymbol{\beta}}}}

\newcommand{\balpha}{{\boldsymbol{\alpha}}}

\newcommand{\Expect}{\mathbb{E}}

\newcommand{\Prob}{\mathrm{P}}

\newcommand{\bX}{\mathbf{X}}
\newcommand{\bx}{\mathbf{x}}

\newcommand{\bpi}{\boldsymbol{\pi}}

\newcommand{\btheta}{\boldsymbol{\theta}}
\newcommand{\bTheta}{\boldsymbol{\Theta}}
\newcommand{\bthetao}{{\boldsymbol{\theta}^o}}
\newcommand{\bthetahat}{\widehat{\boldsymbol{\theta}}}

\newcommand{\bita}{\boldsymbol{\eta}}

\newcommand{\bitahat}{\widehat{\boldsymbol{\eta}}}

\newcommand{\bB}{\mathbf{B}}

\newcommand{\bO}{\mathbf{O}}

\newcommand{\bzero}{\mathbf{0}}

\newcommand{\bR}{\mathbf{R}}

\newcommand{\bbG}{\mathbb{G}}
\newcommand{\bbL}{\mathbb{L}}

\newcommand{\bbP}{\mathbb{P}}
\newcommand{\bbR}{\mathbb{R}}
\newcommand{\dist}{\mathrm{d}}

\newcommand{\mL}{\mathcal{L}}

\newcommand{\Upsilonhat}{\widehat{\Upsilon}}

\newcommand{\sgn}{\mathrm{sgn}}

\usepackage{stackengine}

\DeclareFontFamily{U}{mathx}{\hyphenchar\font45}
\DeclareFontShape{U}{mathx}{m}{n}{
      <5> <6> <7> <8> <9> <10>
      <10.95> <12> <14.4> <17.28> <20.74> <24.88>
      mathx10
      }{}
\DeclareSymbolFont{mathx}{U}{mathx}{m}{n}
\DeclareFontSubstitution{U}{mathx}{m}{n}
\DeclareMathAccent{\widecheck}{0}{mathx}{"71}
\DeclareMathAccent{\wideparen}{0}{mathx}{"75}

\def\titlefull{{Dynamic Classification of Latent Disease Progression with Auxiliary Surrogate Labels}}

\usepackage[]{cleveref}
\usepackage{comment}
\crefname{enumi}{}{}

\newlist{condition}{enumerate}{10}
\setlist[condition]{}
\setlist[condition,2]{}
\crefname{conditioni}{condition}{conditions}
\Crefname{conditioni}{Condition}{Conditions}
\crefname{conditionii}{condition}{conditions}
\Crefname{conditionii}{Condition}{Conditions}
\makeatletter
\addto\extrasenglish{\def\itemautorefname{\@gobble}}
\makeatother
\theoremstyle{definition} 
\newtheorem{theorem}{Theorem}%

\newtheorem*{theorem*}{Theorem}
\newtheorem*{lemma*}{Lemma}

\crefname{appx}{Appendix}{Appendices}
\Crefname{appx}{Appendix}{Appendices}

\usepackage{authblk}

\begin{document}
\def\spacingset#1{\renewcommand{\baselinestretch}%
{#1}\small\normalsize} \spacingset{1}

\title{\Large\bf \titlefull}
\author[1]{Zexi Cai}
\author[2]{Donglin Zeng}
\author[3,4,5,6]{Karen S. Marder}
\author[3,5,6]{Lawrence S. Honig}
\author[1,4,*]{\authorcr Yuanjia Wang}
\affil[1]{Department of Biostatistics, Columbia University, New York, USA}
\affil[2]{Department of Biostatistics, University of Michigan, Ann Arbor, USA}
\affil[3]{Department of Neurology, Columbia University Medical Center, New York, USA}
\affil[4]{Department of Psychiatry, Columbia University Medical Center, New York, USA}
\affil[5]{The Taub Institute for Alzheimer's Disease and the Aging Brain, Columbia University Medical Center, New York, USA}
\affil[6]{Gertrude H. Sergievsky Center, Columbia University Medical Center, New York, USA}
\affil[*]{Email: yw2016@cumc.columbia.edu.}
\date{}
\maketitle

\begin{abstract}
Disease progression prediction based on patients' evolving health information is challenging when true disease states are unknown due to diagnostic capabilities or high costs. For example, the absence of gold-standard neurological diagnoses hinders distinguishing Alzheimer's disease (AD) from related conditions such as AD-related dementias (ADRDs), including Lewy body dementia (LBD). Combining temporally dependent surrogate labels and health markers may improve disease prediction. However, existing literature models informative surrogate labels and observed variables that reflect the underlying states using purely generative approaches, limiting the ability to predict future states. We propose integrating the conventional hidden Markov model as a generative model with a time-varying discriminative classification model to simultaneously handle potentially misspecified surrogate labels and incorporate important markers of disease progression. We develop an adaptive forward-backward algorithm with subjective labels for estimation, and utilize the modified posterior and Viterbi algorithms to predict the progression of future states or new patients based on objective markers only. Importantly, the adaptation eliminates the need to model the marginal distribution of longitudinal markers, a requirement in traditional algorithms. 
Asymptotic properties are established, and significant improvement with finite samples is demonstrated via simulation studies. Analysis of the neuropathological dataset of the National Alzheimer's Coordinating Center (NACC) shows much improved accuracy in distinguishing LBD from AD.
\end{abstract}

\noindent%
\vfill

\newpage
\spacingset{1.8}

\maketitle
\doublespacing
\allowdisplaybreaks

\section{Introduction}
Dynamic classification aims to construct decisions rules from data that evolves over time, accounting for temporal dependencies and changes in patterns. 
However, current methods for this goal often depend on the knowledge of true labels \citep[e.g.,][]{hu2023review}, which can be expensive or infeasible to obtain in real-world applications. For example, the gold-standard neuropathological diagnoses of Alzheimer's disease (AD) and Lewy body dementia (LBD) are only available through postmortem autopsy, making them inaccessible during patients' lifetimes. 
In a large nationwide study of Alzheimer's and related disorders (ADRDs) coordinated by the National Alzheimer's Coordinating Center (NACC) \citep{besser2018version}, only clinical diagnoses are provided as surrogate labels by trained neurologists based on observed symptoms during clinical evaluations. These surrogate labels, while valuable, do not reflect the underlying biological process, suffer from variability in symptom presentations among patients, and, thus, cannot be regarded as gold-standard labels for predicting disease states.

To mitigate the information loss due to {insufficient} 
gold-standard labels, \cite{wang2013auxiliary} considered borrowing information from surrogate disease-informative markers that can partially represent the unobserved disease labels. This approach, however, can only handle cross-sectional data. The hidden Markov model (HMM), a generative model, is a popular and useful technique to handle sequence data with discrete, regular measurements where some or all true labels are missing, but surrogate labels are available; a holistic review can be found in \cite{rabiner1989tutorial}. Inferring the true state status from surrogate labels in longitudinal settings is more complex since 
the collected data often exhibits irregularity. This irregularity may arise from varying observation frequencies among subjects or different measurement time intervals for the same subject. 

Traditional HMMs have been generalized to continuous time to accommodate irregularity and have found wide applications in disease progression modeling  \citep{jackson2002hidden,jackson2003multistate,benoit2016hidden,rubin2017joint,zhou2020continuous,liu2021bayesian}. These methods typically consider the emission of surrogate labels as univariate output conditional on the true latent states and included additional variables in the transition of Markov chains. However, in practice, some variables are direct manifestations of the underlying true health states rather than affecting the transition between health states. For example, in the NACC study, there are longitudinal disease markers, including objective markers and clinical evaluations of cognition and mental health, that directly measure the underlying disease states and, hence, should not be used to model the transition.

To handle multidimensional emissions, generative models are proposed by imposing distributional and/or correlation assumptions on the emitted variables. For example, \cite{song2017hidden} modeled the output variables as a multivariate normal distribution given the latent states, \cite{deruiter2017multivariate} assumed conditional independence among every component of the outputs, and \cite{martino2020multivariate} used multivariate copulas. However, these models may suffer from misspecification and the curse of dimensionality. As the dimension of the emitted variables increases, modeling their joint distribution or correlation structure becomes prohibitive. Moreover, it is often useful to infer new patients' true states or disease progression based on their characteristics and evolving health information. This is a sequential prediction problem, requiring the model to dynamically forecast future disease progression based on all observed historical data. The generative models mentioned above are not fully equipped to handle this goal, as they are not designed to predict future disease progression with accumulating information.

We propose a novel method to address the common challenge of dynamical forecasting of future progression in the absence of true labels over time. The goal is to learn the latent dynamics with surrogate labels and objective markers in the model fitting stage, and rely only on objective markers in the prediction stage. Our method combines the strengths of both generative and discriminative modeling. In the generative model component, we treat the underlying progression as a continuous process to accommodate irregular measurement intervals and introduce a continuous-time HMM that links the surrogate labels with true states. Meanwhile, the discriminative component models the true states given observed variables without modeling the complex joint distribution of variables over time. In fact, the discriminative component not only provides an interpretable sequential classification model but also improves the estimation of the generative counterpart.
We introduce a pseudo-Expectation Maximization (EM) algorithm to handle latent states and introduce regularization in the presence of high-dimensional variables. Our approach leverages an adaptive forward-backward algorithm and Viterbi algorithm to dynamically predict future disease progression based only on objective variables as new data accumulates, which is drastically different from existing HMM models that still rely on the surrogate labels. We study the theoretical properties of the proposed method, including the consistency and the asymptotic normality of the parameters in the model. Extensive simulation studies demonstrate that our method outperforms competing alternatives, and sensitivity analyses show its robustness. We apply our approach to the NACC Neuropathological data, and we find substantial improvements in the accuracy of LBD diagnosis compared to the alternatives.

To the best of our knowledge, the proposed method is the first to integrate the power of both generative and discriminative models for dynamic classification. Our approach offers several advantages. First, it is time-sensitive, incorporating longitudinal surrogate labels and evolving variables as they accumulate over time. Second, it does not require modeling the joint distribution or correlation between the observed variables and, hence, is applicable to high-dimension. Third, by separating the discriminative model from the generative model that uses surrogate labels, the learned discriminative model can be used independently based on accessible and objective variables for future prediction. Finally, under high-dimension, variable selection is easily achieved through appropriate penalization to select the minimum set of markers that can reliably predict the underlying true disease state from a potentially large candidate pool. This facilitates rapid decision-making and minimizes patient burden in clinical settings.

The remainder of this article is structured as follows. The next section discusses the motivating example. \Cref{sec:methodology} introduces our general framework and presents a {pseudo-}EM algorithm to handle missing true disease states. We develop two algorithms that use only the objective variables for predictions on future subjects. In \Cref{sec:asymptotic}, we study the theoretical properties of the proposed methodology. In \Cref{sec:simulation}, we present extensive simulation studies to compare with competing alternatives and conduct sensitivity analyses to examine the method's robustness under various scenarios. We apply the proposed method in \Cref{sec:realdata} to the NACC dataset. We include concluding remarks in \Cref{sec:discussion}, and defer all proofs and supporting information to the Supplementary Materials.

\subsection{The NACC Data}

\begin{figure}[!htpb]
    \centering 
    \includegraphics[width=0.9\textwidth]{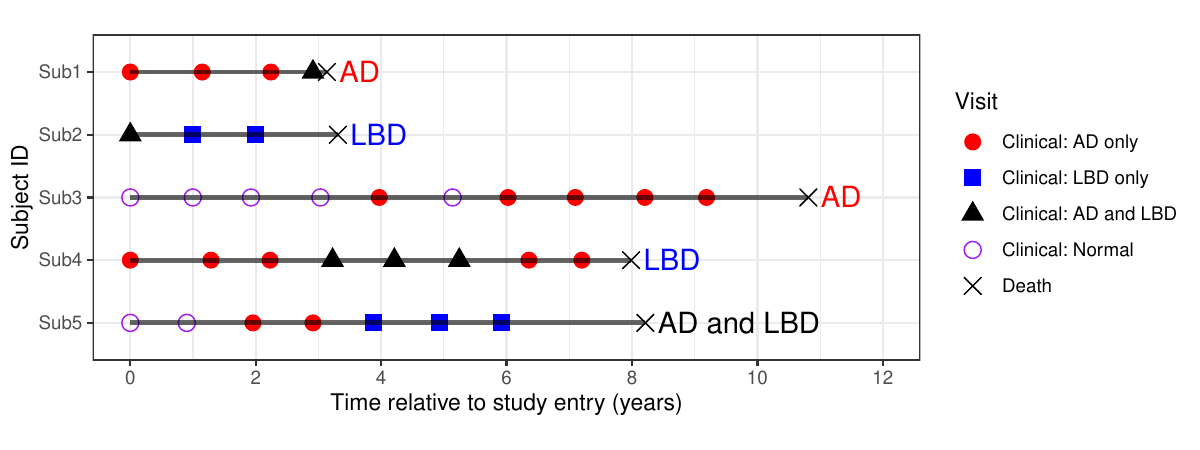}
    \caption{Examples of clinical and gold-standard diagnoses of AD and LBD in the NACC data. Clinical and gold-standard diagnoses are represented by symbols and texts, respectively; AD, LBD, and AD\&LBD diagnoses are represented by red, blue, and black colors.}
    \label{fig:realdata}
\end{figure}

AD accounts for 60\%--80\% of more than 50 million dementia cases worldwide, while LBD is the next most common form of dementia. Pathologically, AD is characterized by the accumulation of amyloid plaques and neurofibrillary tangles, while LBD is characterized by alpha-synuclein deposits in Lewy bodies. Although biologically distinct, these two progressive neurodegenerative disorders share many similarities in early clinical presentation, making it difficult to discern {{AD from LBD or their mixed-type dementia}} in the early stages \citep{weiner1996alzheimer,guerreiro2016genome}. Current diagnosis of AD and LBD relies on clinical examination and {various fluid and imaging} biomarkers {\citep{scott2022fluid,jain2023atn}, which are neither always obtained {due to their invasive or expensive nature}, nor are always perfect.

In the NACC Neuropathology data, {only the neuropathological diagnosis confirmed with autopsy can be treated as the gold-standard true labels}.} 
\Cref{fig:realdata} shows five example subjects from the NACC study, highlighting several challenges of clinical diagnosis. 
For example, Subjects 1 and 4 show that
{one disease can be misdiagnosed as the mixed type or the other type at some visits}. Subjects 3 and 4 illustrate inconsistencies between clinical diagnosis and the gold-standard diagnosis, and a subject can be diagnosed as ``Normal'' after a previous ``AD-only'' diagnosis (Subject 3). {These discrepancies often occur in mild cases where symptoms are subtle.} Misdiagnoses can lead to missed opportunities for early intervention{, suboptimal treatments} 
during later stages of disease progression{, or the inclusion of patients without the correct disease pathologies in clinical trials.} 
Therefore, it is crucial to improve the diagnosis accuracy of the neuropathological stage (latent true disease state) by identifying subtle prodromal measures (objective markers) and leveraging clinical diagnosis (surrogate label).

\section{Methodology}\label{sec:methodology}
\subsection{The Generative-Discriminative Framework}

{Let $\{D(t): t\in[0, \infty)\}$ denote the unobserved true disease process with $D(t)\in\mathcal{D}=\{0,1, \cdots, K\}$. 
{Its progression 
is modeled as a continuous-time 
{Markov chain}, defined by an initial state probability vector $\bpi = (\pi_{0}, \pi_{1} \ldots, \pi_{K})^\top$ with $\sum_{k=0}^{K} \pi_k = 1$, and an instantaneous transition intensity matrix $\bR = (\rho_{kd})_{k,d=0,\ldots,K}$ with non-diagonal elements $\rho_{kd} \geq 0$ and diagonal elements $\rho_{kk} = -\sum_{d\neq k} \rho_{kd}$ for $k=0,\ldots,K$. 
Under the Markov property, for any $s, t \in [0, \infty)$ and $d \in \mathcal{D}$,
$
    \Prob(D(t+s)=d \mid \mathfrak{D}_t) 
    = \Prob(D(t+s)=d \mid D(t)),
$
where $\mathfrak{D}_t=\sigma\{D(s): s \in [0,t]\}$ is the sigma-algebra generated by the disease process up to time $t$. The transition probability, $\Prob(D(s) = d_{s} \mid D(t) = d_{t} )$ for any $s > t$, is the $(d_{t}, d_{s})$-th entry of the matrix exponential, $\exp\left\{(s - t){\bR} \right\}$. It follows that for any $0 = t_1 < t_2 < \ldots < t_m \in [0, \infty)$ with $d_1, \ldots, d_m \in \mathcal{D}$ being any realizations of the disease processes, we have
$
    \Prob(D(t_{1:m})=d_{1:m}) 
    = \Prob(D(t_{1})=d_{1}) 
    \prod_{j=2}^{m} \Prob(D(t_{j})=d_{j}\mid D(t_{j-1})=d_{j-1}),
$
where the subscript ${j_1:j_2}$ denotes all observations indexed by $({j_1},\ldots, {j_2})$.
}

In many applications, although $D(t)$ is unknown, some imperfect surrogate labels, e.g., expert clinical diagnoses, are obtained at multiple longitudinal follow-up times. We consider these surrogate labels as realizations of a stochastic process $\{Z(t):t\in[0, \infty)\}$. Let $\{\bX(t): t\in[0, \infty)\}$ denote a $p$-variate stochastic process of time-varying measures informative of underlying disease states (e.g., cognitive tests, spinal fluid, neuroimaging biomarkers), taking values in $\mathcal{X}\subset\bbR^p$. The joint vector $\left( \bX(t), Z(t) \right)$ then represents the emitted variables from the underlying state $D(t)$. In HMMs, it is common practice to assume temporal conditional independence between observed time points given latent states, i.e., $\Prob(\bX(t_{1:m}), Z(t_{1:m}) \mid D(t_{1:m}) = d_{1:m}) = \prod_{k=1}^{m} \Prob(\bX(t_{k}), Z(t_{k}) \mid D(t_{k}) = d_k)$; see Eq. (13a) of \cite{rabiner1989tutorial}. At the cost of a more complicated discriminative structure presented later, our framework can allow autoregressive HMMs to address residual time dependence in the emitted variables.

As reviewed in Section 1, existing literature often models the conditional distribution $\Prob(\bX(t), Z(t) \mid D(t))$ by imposing stringent distributional assumptions. However, it is difficult to correctly specify this distribution over time, especially for multivariate or even high-dimensional $\bX(t)$. To learn latent states and achieve dynamic prediction, we propose a novel framework that integrates advantages of two mainstream approaches in machine learning, i.e., the generative and discriminative models, as illustrated in \Cref{fig:proposed}. Specifically, to learn $D(t)$, we use a generative approach (e.g., HMM) to incorporate surrogate labels $Z(t)$, the costly or subjective measures with high uncertainty (e.g., clinical diagnosis made by physicians), while we use a classification model to incorporate $\bX(t)$, usually objective or reliably measured data from subjects (e.g., objective biomarkers or granular clinical measures that target specific disease domains).

The model structure, presented in \Cref{fig:proposed}, separates two sources of information: the generative component handles the uncertain and costly labels $Z$, while the discriminative component provides a prediction rule based solely on the objective measures $\bX$. This separation enables learning the latent progression with both $\bX$ and $Z$ in the training, while allows for excluding $Z$ when forecasting disease progression in independent subjects, making predictions reliable and feasible when $Z$ cannot be obtained.

\begin{figure}[!htpb]
\centering
         \begin{tikzpicture}[scale=0.7,transform shape,baseline = (current bounding box.west)]
                \tikzset{
                  main/.style={circle, minimum size = 15mm, thick, draw =black!80, node distance = 8mm and 18mm},
                  connect/.style={-latex, thick},
                  font=\small
                }
                  \node[main] (D1) {$D(t_{1})$};
                  \node[main] (D2) [right=of D1,xshift=-9mm] {$D(t_{2})$};
                  \node[main] (D3) [right=of D2,xshift=5mm] {$D(t_{3})$};
                  \node[main,xshift=-12mm,minimum size = 1mm,fill=white,draw=none] (Ddot) [right=of D3] {\ldots};
                  \node[main,xshift=-12mm] (Dt) [right=of Ddot] {$D(t_{m})$};
                  \node[main,fill=black!10] (X1) [above=6mm of D1] {$\bX(t_{1})$};
                  \node[main,fill=black!10] (X2) [above=6mm of D2] {$\bX(t_{2})$};
                  \node[main,fill=black!10] (X3) [above=6mm of D3] {$\bX(t_{3})$};
                  \node[main,fill=black!10] (Xt) [above=6mm of Dt] {$\bX(t_{m})$};
                  \node[main,fill=black!10] (Z1) [below=6mm of D1] {$Z(t_{1})$};
                  \node[main,fill=black!10] (Z2) [below=6mm of D2] {$Z(t_{2})$};
                  \node[main,fill=black!10] (Z3) [below=6mm of D3] {$Z(t_{3})$};
                  \node[main,fill=black!10] (Zt) [below=6mm of Dt] {$Z(t_{m})$};
                  \path (D1) edge [connect] node[above]{}(D2)
                        (D2) edge [connect] node[above]{}(D3)
                        (D3) edge [connect] (Ddot)
                        (Ddot) edge [connect] (Dt);
                  \path (D1) edge [connect] (X1);
                  \path (D2) edge [connect] (X2);
                  \path (D3) edge [connect] (X3);
                  \path (Dt) edge [connect] (Xt);
                  \path (D1) edge [connect] (Z1);
                  \path (D2) edge [connect] (Z2);
                  \path (D3) edge [connect] (Z3);
                  \path (Dt) edge [connect] (Zt);
                  \node[draw,inner xsep=6pt,inner ysep=12pt+2pt,yshift=-6pt-2pt,dashed,line width=0.5mm,fit={(Z1) (Z2) (Z3) (Zt) (D1) (D2) (D3) (Dt)},label={[label distance=-24pt]-90:{\Large Generative modeling}}, rounded corners] {};
                  \node[draw,inner xsep=6pt,inner ysep=12pt+2pt,yshift=6pt+2pt,solid,fit={(X1) (X2) (X3) (Xt) (D1) (D2) (D3) (Dt)},label={[label distance=-24pt]90:{\Large Discriminative modeling}}, rounded corners] {};
        \end{tikzpicture}
\caption{The proposed modeling approach. White nodes represent latent variables, while grey nodes represent observed data. The generative component usually handles costly or subjective measures, whereas the discriminative component handles objective and reliably measured data.}
\label{fig:proposed}
\end{figure}
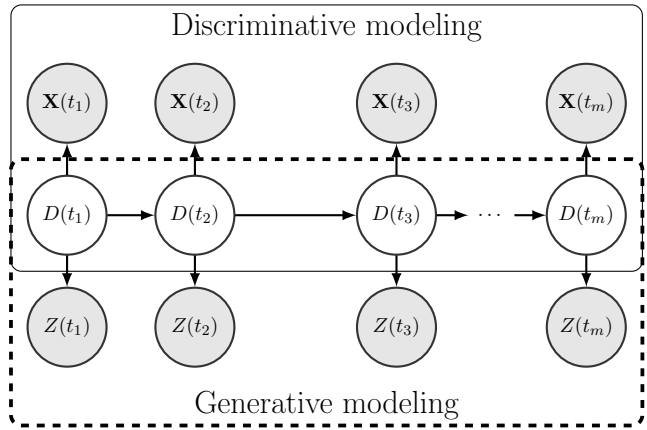

For the generative component, as is common in continuous-time HMMs, a multinomial distribution on $\mathcal{D}$ is used, i.e., we let $e_{dk}=\Prob(Z(t) = k \mid D(t) = d)$ denote the conditional probabilities, and let $\Omega$ denote the emission matrix of $Z$, which contains the parameters $\{e_{dk}\}_{d,k=0,\ldots,K}$. To ensure the identifiability of the parameters, we require that there exists at least one of {the true values of} $\{e_{d1}, \ldots, e_{dK}\}$ that differs between groups $d = 0, \ldots, K$. This assumption is often satisfied in practice; for example, clinicians' misdiagnosis rates for different true disease classes tend to differ. True disease states may be available at some observation times in an application. For example, in our motivating NACC study, AD or LBD state is observed via autopsy confirmation at the last visit, i.e., $Z(t_{m}) = D(t_{m})$ {with probability one}, where $m$ is the total number of observations. For simplicity, we focus on this setting in this article without loss of generality. The estimation procedure remains essentially the same, but more tedious technical details are needed for the proofs under the general setting. 
Moreover, although we present our model with a univariate surrogate label $Z$ under a categorical distribution, we do not restrict the dimension or the distribution of $Z$ as long as its conditional distribution given $D$ can be specified.

For the discriminative modeling, it is common to assume that when the true labels of disease state $D(t)$ are known, the surrogate labels $Z(t)$ do not provide additional information for objective markers $\bX(t)$. For example, changes in disease biomarkers $\bX(t)$ are caused by disease progression, not by receiving which clinical diagnosis. We then propose to reparametrize the emission of $\bX(t)$ using the Bayes rule: 
$
    \Prob(\bX(t) \mid D(t)=k) 
    = \Prob(D(t) = k \mid \bX(t)) \Prob(\bX(t))/
            \Prob(D(t) = k).
$
This reparametrization leverages a generic model for $\Prob(D(t) = k \mid \bX(t))$, requiring only a discriminative approach rather than a complete specification of the conditional distribution of $\bX(t)$ given $D(t)$. Additionally, there is no need to specify the marginal distribution of $\bX(t)$, as it is fully determined by the discriminative model and the generative counterpart. The discriminative model can be any standard model such as logistic regression, probit regression, random forests, and neural networks. In this article, we focus on a time-varying multinomial logistic regression, which is interpretable and effective for modeling multiple classes. Details of the discriminative model are provided in the next section.

\subsection{Pseudo-Expectation Maximization}\label{sec:EM}
In practice, individuals are assessed at discrete times with different frequencies instead of observing the entire stochastic processes. For the same individual, data can be obtained at irregular time intervals. Specifically, for the $i$th subject, $i=1,\ldots,n$, let $T_{ij}$ be the $j$th observation time with $0 = T_{i1} < \cdots < T_{im_i}$, where $m_i$ denotes the number of observations for individual $i$. We assume the distribution of observation times does not involve any parameters of interest. The disease markers are observed as $\bX_{ij} \coloneqq \bX_i(T_{ij})$, and the clinical diagnoses are observed as $Z_{ij} \coloneqq Z_i(T_{ij})$. The observed data from $n$ independent and identically distributed (i.i.d.) subjects are denoted by $\{(m_i, \bX_{ij}, Z_{ij}, T_{ij})\}_{i = 1,...,n, j = 1, \ldots, m_i}${, which can be also viewed as $n$ i.i.d.~realizations from $\bO = (m, \bX_{1:m}, Z_{1:m}, T_{1:m})$}. 
Let 
$\btheta = \allowbreak (\Upsilon, \allowbreak \Lambda)^\top$ 
denote all parameters, where $\Upsilon = \allowbreak(\mathrm{param}(\bpi)^\top, \allowbreak \mathrm{param}(\bR)^\top, \allowbreak \mathrm{param}(\Omega)^\top)^\top$ parameterizes the continuous-time HMM with emissions of $Z$, 
$\Lambda$ parameterizes the discriminative model for $\Prob(D(t) = k \mid \bX(t))$, 
and $\mathrm{param}(\vartheta)$ denotes the free parameters of $\vartheta$. 
Treating unobserved disease states $D_{ij} \coloneqq D_i(T_{ij})$ as missing data, the log-likelihood of the complete data for individual $i$ is
\begin{align*}
    &\ \ell_i(\btheta; \bO_i) \\
    \propto &\ \sum_{j=1}^{m_i-1} \sum_{k=0}^{K} I(D_{ij}=k) \log e_{k,Z_{ij}} 
        + \sum_{j=1}^{m_i} \sum_{k=0}^{K} I(D_{ij}=k) \log \frac{\Prob_{ij}(k\mid \bX_{ij})}
                                                                {\Prob_{ij}(k)} 
        + \sum_{k=0}^{K} I(D_{i1}=k) \log \pi_{k} \\
    &\qquad + \sum_{j=2}^{m_i-1} \sum_{k=0}^{K} \sum_{l=0}^{K} I(D_{i,j-1}=k, D_{ij}=l) \log q_{ijkl} 
        + \sum_{k=0}^{K} I(D_{i,m_i-1}=k) \log q_{im_i k D_{im_i}},
\end{align*}
where {$\Prob_{ij}(k) = \Prob(D_i(T_{ij})=k)$, and} $q_{ijkl} = \Prob(D_{ij} = l \mid D_{i,j-1} = k)$ for $k,l \in \mathcal{D}$. The normalization term $\Prob_{ij}(k)$ can {then} be derived from the discriminative model by $\Prob_{ij}(k) = \int_{\bx} \Prob_{ij}(k\mid \bx) \Prob(\bX_{ij}=\bx) d\bx$. 
We approximate $\Prob_{ij}(k)$ by the Nadaraya-Watson type estimator \mbox{\citep{stute1986almost}}
\begin{equation*}    
    \widehat{\Prob}_{n,ij}(k) 
    = \frac{\sum_{i'=1}^{n} \sum_{j'=1}^{m_{i'}} \Prob_{ij}(k \mid \bX_{i'j'}) \kappa(\frac{T_{ij}-T_{i'j'}}{a_n})}
           {\sum_{i'=1}^{n} \sum_{j'=1}^{m_{i'}} \kappa(\frac{T_{ij}-T_{i'j'}}{a_n})},
\end{equation*}
where $\kappa(\cdot)$ is the Epanechnikov kernel $\kappa(t) = \frac{3}{4} (1 - t^2) I(\abs{t} \leq 1)$, and {$a_n$ is the bandwidth.}%

We now describe the following {pseudo-}EM algorithm for estimation. Denote the parameters obtained at the $m$-th step as $\bthetahat^{(m)}$. The E-step is derived in a similar spirit as the forward-backward algorithm of HMM but with an important and novel adaptation. Specifically, taking the expectation of $\ell_i$ conditioning on the observed data and replacing $\Prob_{ij}(k)$ by $\widehat{\Prob}_{n,ij}(k)$ gives the pseudo E-step
\begin{align}\label{eqn:Qfun}
    Q_i(\btheta; \bthetahat^{(m)}) 
    &= \sum_{j=1}^{m_i-1} \sum_{k=0}^{K} \sum_{l=0}^{K} I(Z_{ij}=l) \widehat{\gamma}_{ijk}^{(m)} \log e_{kl} 
        + \sum_{j=1}^{m_i} \sum_{k=0}^{K} \widehat{\gamma}_{ijk}^{(m)} \log \frac{\Prob_{ij}(k\mid \bX_{ij})}
                                                                                 {\widehat{\Prob}_{n,ij}(k)} 
        + \sum_{k=0}^{K} \widehat{\gamma}_{i1k}^{(m)} \log \pi_{k} \nonumber\\
    &\qquad 
        + \sum_{j=2}^{m_i-1} \sum_{k=0}^{K} \sum_{l=0}^{K} \widehat{\xi}_{ijkl}^{(m)} \log q_{ijkl} 
        + \sum_{k=0}^{K} \sum_{l=0}^{K} I(D_{i,m_i}=l) \widehat{\gamma}_{i,m_i-1,k}^{(m)} \log q_{m_ikl},
\end{align}
where $\widehat{\gamma}^{(m)}_{ijk}= \Prob(D_{ij}=k \mid \bO_i; \bthetahat^{(m)} )$ and $\widehat{\xi}_{ijkl}^{(m)} = \Prob(D_{i,j-1}=k, D_{ij}=l \mid \bO_i; \bthetahat^{(m)} )$. Define the \textit{adaptive} forward-backward probabilities as
\begin{equation*}
    A_{ij}(d; \btheta ) 
    = \Prob(Z_{i,1:j}, \bX_{i,1:j}, D_{ij} = d ; \btheta ) 
        \left/ \prod_{l=1}^{j} \Prob(\bX_{il} ; \btheta )\right.,
\end{equation*}
and
\begin{equation*}
    B_{ij}(d; \btheta ) 
    = \Prob(Z_{i,j+1:m_i}, \bX_{i,j+1:m_i} \mid D_{ij} = d; \btheta ) 
        \left/ \prod_{l=j+1}^{m_i} \Prob(\bX_{il}; \btheta )\right.,
\end{equation*}
then 
\begin{equation*}
    \gamma_{ijk} 
    = \frac{A_{ij}(k) B_{ij}(k)}
           {\sum_{k=0}^{K} A_{i,m_i}(k)}, 
    \qquad 
    \xi_{ijkl} 
    = A_{i,j-1}(k) B_{ij}(l) 
        e_{l,Z_{ij}} q_{ijkl} 
        \frac{\Prob_{ij}(l \mid \bX_{ij})}
            {\widehat{\Prob}_{n,ij}(l)} 
        \left/\sum_{k=0}^{K} A_{i,m_i}(k)\right.,
\end{equation*}
where we omit the dependence on $\btheta$ for brevity when there is no ambiguity. These modifications are non-trivial: the original form of forward-backward probabilities does not include the denominators, which leaves the marginal distribution $\Prob(X_{il};\btheta)$ in the recursive formula of $A$, $B$, and $\xi$; in contrast, we introduce such an offsetting term as an important adaptation so that no marginal distribution of $\bX_{ij}$ is required to obtain $\gamma_{ijk}$ or $\xi_{ijkl}$, based on which they can be simply set to any constant $c$ (e.g., $c=1$).  This is highly desirable since modeling the distribution of high-dimensional variables $\bX_{ij}$ over time is difficult. The recursion formula for the adaptive forward-backward probabilities and the detailed derivations of the above results are deferred to Appendix A.1.

The M-step is to maximize the function $Q(\btheta; \bthetahat^{(m)}) = n^{-1} \sum_{i=1}^{n} Q_i(\btheta; \bthetahat^{(m)})$ with respect to $(\Upsilon, \Lambda)$ using coordinate ascent. For $\Upsilon$, the first step is to maximize it with respect to the initial state probability $\bpi$ and the emission matrix $\Omega$ using the closed-form updates
$
    \widehat{\pi}_{k}^{(m+1)} 
    = n^{-1} \sum_{i=1}^{n} \widehat{\gamma}_{i1k}^{(m)}   
$
for $k = 0, \ldots, K$, and
$
    \widehat{e}_{kl}^{(m+1)} 
    = \sum_{i=1}^{n} \sum_{j=1}^{m_i} I(Z_{ij}=l) \widehat{\gamma}_{ijk}^{(m)}
           / \sum_{i=1}^{n} \sum_{j=1}^{m_i} \widehat{\gamma}_{ijk}^{(m)}
$
for $k,l = 0, \ldots, K$. 
Next, optimizing $Q$ with respect to $\bR$ corresponds to maximizing 
$\sum_{j=2}^{m_i-1} \sum_{k=0}^{K} \sum_{l=0}^{K} \widehat{\xi}_{ijkl}^{(m)} \log q_{ijkl} + \sum_{k=0}^{K} \sum_{l=0}^{K} I(D_{i,m_i}=l) \widehat{\gamma}_{i,m_i-1,k}^{(m)} \log q_{m_ikl}$, and we use local maximization tools {such as the Nelder-Mead method} with multiple starting points since no closed-form solution exists for the transition parameters. A practical starting point is the transition intensity matrix fitted by a continuous-time HMM model to the data without considering $\bX$. The optimizer is denoted as $\widehat{\bR}^{(m+1)}$.

The optimization with respect to $\Lambda$ depends on the specific choice of the discriminative model. In this article, we consider the multinomial logistic model with a time-varying intercept with 
\begin{equation}\label{eqn:tvlogistic}
    \Prob(D(t) = k \mid \bX(t))
    = \frac{\exp(\alpha_{k}(t) + \bbeta_k^\top \bX(t))I(k\neq 0) + I(k=0)}{1+\sum_{d=1}^{K} \exp(\alpha_{d}(t) + \bbeta_d^\top \bX(t))}, 
    \quad k=0,\ldots,K.
\end{equation}
Here, $\alpha_{k}(t)$ is a nonparametric time-varying intercept and $\bbeta_k = (\beta_{k1}, \ldots, \beta_{kp})^\top \in \bbR^p$ is a vector of coefficients. Including a time-varying intercept is important to capture the change in the prevalence of each disease state over time. Specifically, recall the conditional logistic regression used to analyze stratified matched groups. Due to stratification, group-specific intercepts are used to account for the different frequencies in the outcome. The time $T$ here becomes a natural stratification factor as the prevalence of each disease class also varies over time due to progression. With \Cref{eqn:tvlogistic}, $\Lambda = \allowbreak (\alpha_1(\cdot), \allowbreak \ldots, \allowbreak \alpha_K(\cdot), \allowbreak \bbeta_1^\top, \allowbreak \ldots, \allowbreak \bbeta_K^\top)^\top$. For ease of notation, we let $\balpha(\cdot) = (\alpha_1(\cdot), \ldots, \alpha_K(\cdot))^\top$ and also reorganize the vector $(\bbeta_1^\top, \ldots, \bbeta_K^\top)^\top$ as $\bbeta = (\bbeta_{(1)}^\top, \ldots, \bbeta_{(p)}^\top)^\top$, where $\bbeta_{(u)} = (\beta_{1u}, \ldots, \beta_{Ku})^\top \in \bbR^{K}$, $u=1,\ldots,p$.

This discriminative model is a parsimonious choice for longitudinal classification that can be easily extended to more complex scenarios. For example, the linear form $\bbeta_k^\top \bX(t)$ can be extended to handle time-varying effects, $\bbeta^\top_k(t) \bX(t)$; if interactions are desired, additive models $\sum_{j=1}^{p} f_{kj}(X_j(t))$ or other nonparametric functions $f_k(\bX(t))$ with random forests or neural networks can be utilized. However, these models are out of the scope of this paper, and we leave the investigation as future work.

The time-varying intercepts are modeled as smooth functions and approximated using B-splines as $\alpha_k(t)=\bita_k^\top \bB_{r}(t)$, where $\bB_{r}(\cdot)$ is a set of known B-spline basis functions of order $r$, and $\bita_k$ is an unknown coefficient. The equidistant knots are set as $0 = t_0 < t_1 < \ldots < t_{J_n} < t_{J_n+1} = t^*$, where $J_n$ denotes the number of interior knots that depends on the sample size, and $t^*$ is an upper bound of the observational time $T$. As suggested by the large sample properties presented in \Cref{sec:asymptotic}, we use cubic splines with $\lceil n^{1/9} \rceil$ interior knots in the simulation and real data analysis. Based on the spline approximation, we have
\begin{equation*}
    \Prob_{ij}(k\mid \bX_{ij})
    \approx \frac{\exp(\bita_k^\top \bB_{r}(T_{ij}) + \bbeta_k^\top \bX_{ij}) I(k\neq 0) + I(k=0)}
                  {1+\sum_{d=1}^{K} \exp(\bita_d^\top \bB_{r}(T_{ij}) + \bbeta_d^\top \bX_{ij})},
    \qquad k=0,\ldots,K.
\end{equation*}

Optimization of $\Lambda$ now corresponds to maximizing $\sum_{j=1}^{m_i} \sum_{k=0}^{K} \widehat{\gamma}_{ijk}^{(m)} \log \{\Prob_{ij}(k\mid \bX_{ij}) / \widehat{\Prob}_{n,ij}(k)\}$ with respect to the parameters $\bita$ and $\bbeta$. While gradient ascent can be used to find the optimizer, we introduce a novel numerical trick that significantly accelerates and stabilizes the computation by leveraging {a} dual parametrization. In addition to the Nadaraya-Watson type estimator, another way to obtain $\Prob_{ij}(k)$ is by using the transition matrix and initial state probabilities from the continuous-time HMM: $\Prob_{ij}(k) = \sum_{l=0}^{K} \Prob(D_{ij} = k \mid D_{i0} = l) \Prob(D_{i0} = l)$, which leads to $\widehat{\Prob}_{n,ij}(k) = \sum_{l=0}^{K} \pi_{l} \exp(T_{ij}{R})^{(l,k)}$. At the convergence of the EM algorithm, using either set of parameters from the dual representation would yield nearly identical $\widehat{\Prob}_{n,ij}(k)$, as both expressions estimate the same population quantity. Our key adaptation is to substitute the denominator $\widehat{\Prob}_{n,ij}(k)$ by the continuous-time HMM representation with $\widehat{\bpi}^{(m+1)}$ and $\widehat{\bR}^{(m+1)}$ in each iteration of the M-step. This substitution transforms the remaining part to be optimized into the objective function of a weighted logistic regression, greatly simplifying the computation and speeding up convergence. Denote the estimates as $\bitahat_k^{(m+1)}$ and $\bbetahat_k^{(m+1)}$. The time-varying intercept is then estimated by $\alphahat_k^{(m+1)}(t) = \bitahat_k^{(m+1)\top} \bB_{r}(t)$.

The proposed algorithm can also easily incorporate variable selection when the disease marker $X$ is of high dimension. In this case, it is desirable to identify important markers to avoid overfitting and ensure generalizability to new patients. A penalty term is introduced to the M-step,  where sparsity is encouraged. Since the number of disease states may be more than two, 
we use group-wise penalties. In particular, in the M-step, we maximize the objective function with a penalty on the multinomial logistic regression parameters, i.e., 
$
    Q(\btheta; \bthetahat^{(m)}) 
    - \sum_{u=1}^{p} \mathcal{P}_{\lambda_u}(\bbeta_{(u)}),
$
where $\mathcal{P}(\cdot)$ is a generic penalty function, and $\lambda_u \geq 0$ is a tuning parameter associated with the $u$th coefficient; we also suppress its dependency on $n$ for brevity. {Examples include 
$\mathcal{P}_{\lambda_u}(\bbeta_{(u)}) = \lambda \|\bbetacheck_{(u)}\|_2^{-1} \|\bbeta_{(u)}\|_2$ for adaptive group lasso, where $\bbetacheck_{(u)}$ is some $n^{1/2}$-consistent estimator of $\bbeta^o_{(u)}$. Alternative penalizations such as the group smoothly clipped absolute deviation penalty and group minimax concave penalty can also be employed.} 
Furthermore, we choose the tuning parameters via the 5-fold cross-validation with the one standard error rule of cross-validation errors {(recall that the associated objective function resembles that of a weighted logistic regression when the normalization term is fixed using the hidden Markov parameterization)}.

Iterating the E- and M-steps of the {pseudo-}EM algorithm until {some convergence criteria are met} gives {a sieve MLE $\bthetahat = \allowbreak (\Upsilonhat, \allowbreak \alphahat_1(\cdot), \allowbreak \ldots, \allowbreak \alphahat_K(\cdot), \allowbreak \bbetahat_1^\top, \allowbreak \ldots, \allowbreak \bbetahat_K^\top)^\top$ with $\alphahat_k(\cdot) = \bitahat_k^\top \bB_{r}(\cdot)$, where we omit in the subscript its dependence on the sample size $n$.} 
In the appendix, we justify our proposed method by investigating its large-sample properties. Specifically, we show 
the identifiability of the model parameters, 
prove the proposed estimator is consistent, and establish the oracle properties of variable selection.

\subsection{Prediction of Disease Progression in Future Subjects}\label{sec:prediction}
The obtained discriminative model enables one to make predictions without knowing the surrogate labels $Z$, but only using future subjects' objective disease markers $\bX$. We present two algorithms analogous to the posterior probability prediction and the Viterbi decoding in HMMs. Assume the $i$th future subject's $j$th visit occurs at time $T_{ij}$ and the corresponding disease markers are $\bX_{ij}$, then the predicted probability in class $k$ is
\begin{equation*}
    \Prob(D_{ij} = k \mid \bX_{ij}; \bthetahat) 
    = \frac{\exp(\alphahat_{k}(T_{ij}) + \bbetahat_k^\top \bX_{ij}) I(k\neq 0) + I(k=0)}
           {1 + \sum_{d=1}^{K} \exp(\alphahat_{d}(T_{ij}) + \bbetahat_d^\top \bX_{ij})},
\end{equation*}
and the posterior probability prediction outputs $k^* = \arg\max_k \Prob(D_{ij} = k \mid \bX_{ij}; \bthetahat)$.

In contrast to finding the most likely state specifically at each time point, Viterbi's decoding algorithm gives the most probable sequence given the transition structure and the observations accrued up to the current time. The corresponding Viterbi's algorithm adapted to our proposed method proceeds as follows (with detailed derivations presented in Appendix A.2): we first define
\begin{equation*}
    \delta_{ij}(d) = 
    \max_{D_{i1},\ldots,D_{i,j-1}} 
    \frac{\Prob(D_{ij}=d, D_{i1}, \ldots, D_{i,j-1}, \bX_{ij}, \bX_{i1}, \ldots, \bX_{i,j-1}; \bthetahat)}
         {\prod_{l=1}^{j} \Prob(\bX_{il}; \bthetahat)},
\end{equation*}
{which quantifies the \textit{adaptive} probability of the most probable sequence with observed $\bX_{i,1:j}$, }then starting with
\begin{equation*}
    \delta_{i1}(d) = 
    \frac{\Prob(D_{i1}=d, \bX_{i1}; \bthetahat)}
         {\Prob(\bX_{i1}; \bthetahat)} 
    = \Prob(D_{i1}=d \mid \bX_{i1}; \bthetahat),
\qquad
    \psi_{i1}(d) = -1,
\end{equation*}
we have the recursion formula as
\[
    \delta_{i,j+1}(d) =
    \frac{\Prob(D_{i,j+1}=d \mid \bX_{i,j+1}; \bthetahat)}
         {\Prob(D_{i,j+1}=d; \bthetahat)} 
    \max_{D_{ij}} \left[\Prob(D_{i,j+1}=d \mid D_{ij}; \bthetahat) \delta_{ij}(D_{ij})\right],
\]
\[
    \psi_{i,j+1}(d) = 
    \arg\max_{D_{ij}} \left[\Prob(D_{i,j+1}=d \mid D_{ij}; \bthetahat) \delta_{ij}(D_{ij})\right],
\]
for $d=0,\ldots,K$. 
At termination, we have the probability of the most probable sequence as $P_i^* = \max_{d} \delta_{im_i}(d)$, and the most probable sequence is obtained by backtracing via $D_{im_i}^* = \arg\max_{d} \delta_{im_i}(d)$ and $D_{ij}^* = \psi_{i,j+1}(D_{i,j+1}^*)$, $j=1,\ldots,m_i-1$. Similarly, the modification here is important as it eliminates the need to model the marginal distribution $\Prob(X_{il};\btheta)$ in the recursive formula. The most probable sequence preserves the ordering of the progression and can be updated adaptively when new observations become available, which can assist with the early detection of disease progression in practice. Moreover, the Viterbi algorithm can be used to forecast disease status at a sequence of future time points, which is useful for disease treatment and management.

\section{Large Sample Properties}\label{sec:asymptotic} 
The parameter space of $\btheta = (\Upsilon, \balpha, \bbeta)^\top$ is denoted as $\bTheta = \mathcal{Y} \times \mathcal{A} \times \mathcal{B}$, and is a subset of $\bbR^{\mathrm{card}(\Upsilon)} \times \{C^s[0,t^*]\}^{K} \times \bbR^{K\times p}$, where $\{C^s[0,t^*]\}^{K}$ is the product of $K$ functional spaces of continuous differentiable functions of order $s$ on $[0,t^*]$. The sieve space $\bTheta_n = \mathcal{Y} \times \mathcal{A}_n \times \mathcal{B}$ is a subset of $\bbR^{\mathrm{card}(\Upsilon)} \times \{C_n^r[0,t^*]\}^{K} \times \bbR^{K\times p}$, where
\[
    C_n^r[0,t^*] 
    = \left\{\bita_n^\top \bB_{r}(\cdot): \bita_n \in \Gamma, \Gamma \subset \bbR^{J_n + r},
    \sum_{i=1}^{J_n+r} \abs{\eta_i} \text{ is bounded}
    \right\}
\] 
denotes the class of functions that can be represented by splines basis functions of order $r$ with $J_n$ interior knots on $[0,t^*]$. {Let $\btheta^o = \allowbreak (\Upsilon^o, \allowbreak \alpha_1^o(\cdot), \allowbreak \ldots, \allowbreak \alpha_K^o(\cdot), \allowbreak \bbeta_1^{o\top}, \allowbreak \ldots, \allowbreak \bbeta_K^{o\top})^\top$ denote all true parameters. Similarly,  let $\btheta_n^o = \allowbreak (\Upsilon^o, \allowbreak \alpha_{n1}^{o}(\cdot), \allowbreak \ldots, \allowbreak \alpha_{nK}^{o}(\cdot), \allowbreak \bbeta_1^{o\top}, \allowbreak \ldots, \allowbreak \bbeta_K^{o\top})^\top$ with $\alpha_{nk}^{o}(\cdot) = \bita_{nk}^{o\top} \bB_{r}(\cdot)$  denote the approximate truth of $\btheta$ in the sieve space. {We suppress the dependence of $\bita$ on $n$ in the sequel when the context is clear.} The true parameter of $\bbeta$ is $\bbetao = (\bbeta_{(1)}^{o\top}, \ldots, \bbeta_{(p)}^{o\top})^\top$. %
Without loss of generality, suppose $\|\bbeta^{o}_{(u)}\|_2 \neq 0$ for $u = 1, \ldots, p_0$ and $\|\bbeta^{o}_{(u)}\|_2 = 0$ for $u = p_0 + 1, \ldots, p$, and we define $\bbetao = (\bbeta_{[1]}^{o\top}, \bbeta_{[2]}^{o\top})^{\top}$ with $\bbeta_{[1]}^{o} = (\bbeta^{o\top}_{(1)}, \ldots, \bbeta^{o\top}_{(p_0)})^\top$, and $\bbeta_{[2]}^{o} = (\bbeta^{o\top}_{(p_0+1)}, \ldots, \bbeta^{o\top}_{(p)})^\top$.} In the sequel, we use $c$ and subscripts, $c_1$, $c_2$, \ldots, to denote constants that may differ in different lines.

Under technical conditions specified in Appendix B, \Cref{lem:identifiability} below shows the identifiability of the model.
\begin{theorem}\label{lem:identifiability}%
    Let 
    \begin{align*}
        \mL(\btheta; \bO) 
        =&\ \log 
            \sum_{d_{m}} \cdots \sum_{d_{1}} \prod_{j=1}^{m} 
            \Prob(Z(T_{j}), \bX(T_{j}) \mid D(T_{j})=d_{j}; \btheta) 
            \pi_{d_{1}} \prod_{j=2}^{m} \exp((T_{j}-T_{j-1})\bR)^{(d_{j-1},d_{j})},
    \end{align*}
    where $(d_{j-1}, d_{j}) \in \mathcal{D}^*$ for $j = 2, \ldots, m$, $(d_{j}, Z_{j}) \in \mathcal{E}^*$ for $j = 1, \ldots, m-1$, and $d_{m} = Z_{m}$. Here $\mathcal{D}^*$ denotes the set of all state pairs $(k, l)$ such that a transition from $k$ to $l$ is feasible, and $\mathcal{E}^*$ denotes the set of all state-observation pairs $(k, l)$ such that emission from $k$ to $l$ is feasible except for the last observational time. Suppose Conditions (C1) to (C5) hold, then the model is identifiable in the sense that $\mL(\btheta; \bO) = \mL(\btheta^*; \bO)$ implies $\btheta = \btheta^*$.
\end{theorem}
Its proof is in Appendix C. Moreover, Lemma 1 in the Appendix indicates that the EM algorithm converges to a local maximum of the penalized observed log-likelihood function.
Multiple initial values $\bthetahat^{(0)}$ shall be used to ensure that the global maximum can be achieved. Under the sieve estimation framework, we have $\bthetahat = \arg\max_{\btheta\in\bTheta_n} \widehat{\mL}_n(\btheta)$, where $\widehat{\mL}_n$ is obtained by replacing the quantity $\Prob_{ij}(k)$ with $\widehat{\Prob}_{n,ij}(k)$. In what follows, with a little abuse of notation, we also use $(\Upsilon,\balpha,\bbeta)$ to represent $\btheta$ whenever it is necessary or more convenient to distinguish the different sets of parameters.

In the following theorems, we study the consistency and the rate of convergence of the estimator in the $L_2$-metrics defined by
\begin{align*}
    \dist(\btheta_1, \btheta_2) 
    = \left\{\norm{\Upsilon_1 - \Upsilon_2}_2^2 
        + \int_0^{t^*} \norm{\balpha_1(s) - \balpha_2(s)}_2^2 ds%
        + \norm{\bbeta_1 - \bbeta_2}_2^2
        \right\}^{1/2}.
\end{align*}
Denote for shorthand that %
$\norm{\balpha_1 - \balpha_2}_{T} \coloneqq \{\int_0^{t^*} \norm{\balpha_1(s) - \balpha_2(s)}_2^2 ds\}^{1/2}$. 
Let $\bbP_n$ denote the empirical measure defined by $\bbP_n f = n^{-1} \sum_{i=1}^{n} f(\bO_i)$ and $\bbG_n$ be the empirical process defined by $\bbG_n f = n^{1/2}(\bbP_n - P) f = n^{-1/2} \sum_{i=1}^{n} (f(\bO_i) - \Expect_P f(\bO_i))$. Let $\bbL_n(\Upsilon, \balpha, \bbeta) = \bbP_n \mL(\Upsilon, \balpha, \bbeta) - \sum_{u=1}^{p} \mathcal{P}_{\lambda_u}(\bbeta_{(u)})$ and $\bbL(\Upsilon, \balpha, \bbeta) = P \mL(\Upsilon, \balpha, \bbeta)$, where $\mL$ is defined in \Cref{lem:identifiability}. We use $\widehat{\bbL}_n(\Upsilon, \balpha, \bbeta)$ to denote the (penalized) criterion function when $\Prob_{ij}(k)$ is replaced by $\widehat{\Prob}_{n,ij}(k)$. {Lemmas 2 and 3 in the Appendix establish the complexity of the class of functions formed by the log-likelihood functions and the separability of the criterion function, respectively, based on which the consistency and the rate of convergence of the estimator are shown in \Cref{thm:theta}.}

\begin{theorem}\label{thm:theta}
    Suppose Conditions (C1) to (C10) hold, then
    \begin{enumerate}[label=(\alph*), ref=\thetheorem (\alph*)]
        \item \label[theorem]{thm:thetaConsistency}
        $
            \dist(\bthetahat, \bthetao) = o_p(1);
        $
    
        \item \label[theorem]{thm:thetaRate}
        $\norm{\bbetahat - \bbetao}_2 = O_p(n^{-1/2})$ and
        $
            \dist(\bthetahat, \bthetao) = O_p(n^{-\min(s\nu, (1-\nu)/2)}).
        $
    \end{enumerate}
\end{theorem}
Guided by the regularity conditions, cubic splines can be used for approximating the intercepts with $\lceil n^{1/9} \rceil$ interior knots, and the kernel bandwidth for obtaining $\widehat{\Prob}_{n,ij}$ can be chosen as $\lfloor n^{-1/5} \rfloor$. 
{\Cref{thm:thetaRate} implies that the optimal convergence rate of the estimator is $O_p(n^{-s/(2s+1)})$, which is achieved when $\nu = 1/(2s+1)$, and although the overall convergence rate for $\bthetahat$ is slower than the parametric rate of convergence $n^{1/2}$, the logistic coefficients can still be estimated at the parametric rate. Moreover, it enjoys the oracle properties with the penalty used, as shown in \Cref{thm:oracle}.}
\begin{theorem}\label{thm:oracle}
    Let $\bbetahat = (\bbetahat_{[1]}^{\top}, \bbetahat_{[2]}^{\top})^{\top}$. Suppose $s\nu > 1/2$ and Conditions (C1) to (C10) hold, then we have
    \begin{enumerate}[label=(\alph*), ref=\thetheorem (\alph*)]
        \item \label[theorem]{thm:selectionConsistency}
        Selection consistency: $\Prob(\bbetahat_{[2]} = \bzero) \to 1$;

        \item \label[theorem]{thm:asymptoticNormality}
        Asymptotic normality: $n^{1/2} (I_1^o+\Sigma_n)(\bbetahat_{[1]} - \bbeta^o_{[1]} + (I_1^o+\Sigma_n)^{-1} \bb_n)$ converges in distribution to $N(\bzero, I_1^o)$, where $I_1^o = I_1(\bbeta^o_{[1]})$ is the Fisher information knowing $\bbeta^o_{[2]} = \bzero$, $\Sigma_n = \mathrm{diag}(-\mathcal{P}''_{\lambda_{1}}(\bbeta^o_{(1)}),\ldots,-\mathcal{P}''_{\lambda_{p_0}}(\bbeta^o_{(p_0)}))$ and $\bb_n = (-\mathcal{P}'_{\lambda_{1}}(\bbeta^o_{(1)}) \sgn(\bbeta^o_{(1)}), \ldots, -\mathcal{P}'_{\lambda_{p_0}}(\bbeta^o_{(p_0)}) \sgn(\bbeta^o_{(p_0)}))^\top$. Here the multivariate signum function $\sgn$ is defined as $\sgn(\bbeta) = \bbeta/\norm{\bbeta}_2$.
    \end{enumerate}
\end{theorem}

{The proof of \Cref{lem:identifiability,thm:theta,thm:oracle} is in Appendices C.4 to C.6,, respectively.}

\section{Simulation Studies}\label{sec:simulation}
\subsection{Simulation Design}
We simulate data similar to the NACC study. Specifically, we generate three disease states (AD-only, LBD-only, or AD\&LBD) in addition to the healthy state, with $0$ denoting the healthy state and 1, 2, or 3 denoting each disease state, respectively. The transition intensity matrix mimics the disease progression of AD and LBD: a healthy patient ($D=0$) can stay healthy or have AD-only ($D=1$), LBD-only ($D=2$), or AD\&LBD ($D=3$) later on; furthermore, the diseased state cannot be recovered. %
Details of the HMM parameters are presented in Appendix D.

Given the true disease states, the disease markers are sampled with the following two steps. The first step is to form clusters of candidate markers for each state. We generate a large number of i.i.d.~copies of $\widetilde{\bX}$ and obtain their attached labels $\widetilde{D}$ from the logistic model, and this forms $K=4$ clusters of disease markers. We may discard the time-varying intercept in this step since no time information is involved. Second, after obtaining $D_{ij}$ from the continuous-time HMM model for each $i$ and $j$, we randomly sample one $\widetilde{\bX}$ as $\bX_{ij}$ from the $\widetilde{D}=D_{ij}$-th variable group.

We consider $p=4$, $p=20$, and $p=500$, where for the latter two cases, only the first four markers are associated with the disease status{, and we apply the group adaptive lasso in the estimation}. We consider three scenarios for the marker distributions: the first two scenarios include continuous and a combination of continuous and discrete variables, respectively, and the third one includes a set of correlated variables. Details of the parameters and data generation are also deferred to Appendix D.

The sample size is $n = 200$ and $500$, and the maximum follow-up time is $t_{\mathrm{max}} = 10$. The total number of visits $m_i$ for each individual follows $2$ plus a Poisson distribution with mean $6$ and we randomly place the $m_i$ visit times on the interval $[0,10)$. For each set of simulations, $500$ replications are conducted. For the {pseudo-}EM algorithm, the starting values of the transition intensity and the emission matrices are obtained by fitting an HMM, and the initial logistic parameters are based on logistic regression with the HMM Viterbi-fitted labels. As kernel approximations are introduced in the estimation process, we do not focus on the $Q$-function for the convergence criterion. Instead, to account for different magnitudes and cardinalities of the parameters, we set the stopping criterion as either the maximum number of iterations allowed, specified as $100$, is achieved, or the minimum of the absolute and relative errors is less than $10^{-4}$. As observed in the simulation and real data analysis, the stopping criteria effectively balance the computational efficiency and accuracy, and the algorithm converges to a stable solution. We also define $0/0=0$ in the relative error computation when $p=20$ or $500$.

We compare our proposal ([\textit{Proposed}]) to the following methods: 
[\textit{$D$-known}] an ideal procedure where the true disease labels at all observation times are used as the outcomes in training a time-varying multinomial logistic model; 
[\textit{HMM}] a simple procedure where a continuous-time HMM is fitted on the observed surrogate labels, and the subjects are classified to the most probable class according to the basic forward-backward or Viterbi algorithms using the surrogate labels; 
[\textit{Obs}] a na\"ive procedure where the expert surrogate diagnosis labels are used in training a time-varying multinomial logistic model. 

To evaluate the prediction performance, for each simulation replicate, an independent test set of twice the training sample size is generated. To demonstrate that our proposal can achieve good prediction when patients are still alive (i.e., no autopsy to obtain gold-standard disease labels), 
{we consider a shorter time interval in the test set as} $[0, 7) \subset [0, 10)$, and the number of visits follows $2$ plus a Poisson distribution with a mean of $3$.  For the posterior prediction, we also include the multivariate generalization of the area under the receiver operating characteristic curve (AUC) by considering all four one-versus-rest classifications. For $p=20$ or $p=500$, the average number of correctly identified important disease markers, the average number of incorrectly included noise variables, {and the Matthews correlation coefficient 
are reported as measures of variable selection performance. 

To assess estimation performance, 
we report the average bias and mean squared error over the elements of the vector of each group of parameters. As the true time-varying intercepts $\alpha_{0k}(\cdot)$'s are implicitly determined by the underlying latent disease labels from the generative model, we compare the mean integrated squared difference 
between the $\alphahat_{0k}(\cdot)$'s estimated from \textit{Proposed} 
and the \textit{$D$-known} method, 
defined as 
$
    \Expect( \int_{0}^{{t_\mathrm{max}}} (\bitahat_{k}^{o\top} \bB_{r}(t) - \bitahat_k^{\top} \bB_{r}(t))^2 dt),
$
where $\bitahat_{k}^{o}$'s are the spline coefficients estimated using \textit{$D$-known}. The integral is computed using numerical integration over 1000 equally spaced grids on $[0, t_{\mathrm{max}}]$.

\subsection{Simulation Results}
The prediction accuracy of the pointwise and sequential prediction methods for Scenario I is summarized in \Cref{tab:cov1p4_tab1}, from which it can be seen that the proposed method's accuracy and AUC are comparable to the ``oracle", \textit{$D$-known}, where the difference is generally less than 1\%. %
{In addition, the proposed method provides an accuracy improvement of more than 15\% and an AUC improvement of more than 5\% compared to the other alternatives.} 
Increasing the sample size from 200 to 500 increases the accuracies for all the approaches, but the advantage of the proposed method remains substantial. {The lower biases and mean squared errors of the HMM progression parameters with \textit{Proposed} compared to \textit{HMM} reveal that the inclusion of the discriminative component also improves the estimation of the generative counterpart.} 
{For $p=4$, Table S1 in the appendix details a logistic model estimation with inference results. For $p=20$ and $500$, the logistic model estimation with variable selection performance is summarized in \Cref{tab:cov1p4_tab2}. } The proposed method provides comparable bias and mean squared error for the logistic parameters compared to the \textit{$D$-known} approach, %
and the mean integrated squared difference for the time-varying intercepts tends to zero as the sample size increases. 
Tables S2 and S3 in the Appendix show similar results in other scenarios.

\begin{table}[!htbp]
  \centering
  \caption{Prediction performance and HMM parameter estimation results with variable scenario (I)}
    \resizebox{\textwidth}{!}
    {%
    \begin{tabular}{ccccccccccccccc}
    \toprule 
    \multirow{3}[0]{*}{$p$} & \multirow{3}[0]{*}{$n$} & \multirow{3}[0]{*}{Method} & \multicolumn{6}{c}{Classification Performance} & \multicolumn{6}{c}{HMM Progression} \\\cmidrule(lr){4-9}\cmidrule(lr){10-15}
         &      &      & \multicolumn{2}{c}{Accuracy \%} & \multicolumn{4}{c}{AUC$\times10^2$ (Posterior)} & \multicolumn{3}{c}{Bias$\times10^3$} & \multicolumn{3}{c}{MSE$\times10^3$} \\\cmidrule(lr){4-5}\cmidrule(lr){6-9}\cmidrule(lr){10-12}\cmidrule(lr){13-15}
         &      &      & Posterior & Viterbi & 0    & 1    & 2    & 3    & Tran & Emis & Init & Tran & Emis & Init \\\midrule
    4    & 200  & $D$-known & 77.98 & 89.49 & 94.67 & 95.13 & 92.65 & 94.84 & -    & -    & -    & -    & -    & - \\
         &      & Proposed & 77.78 & 89.13 & 94.58 & 95.05 & 92.56 & 94.79 & 1.76 & 1.27 & 1.7  & 0.93 & 0.5  & 0.7 \\
         &      & HMM  & 49.21 & 55.77 & 75.08 & 58.89 & 64.05 & 57.8 & 13.52 & 2.89 & 15.15 & 5.11 & 0.89 & 4.82 \\
         &      & Obs  & 61.05 & -    & 88.93 & 91.57 & 88.95 & 91.47 & -    & -    & -    & -    & -    & - \\
         & 500  & $D$-known & 78.22 & 89.63 & 94.77 & 95.21 & 92.76 & 94.94 & -    & -    & -    & -    & -    & - \\
         &      & Proposed & 78.15 & 89.48 & 94.74 & 95.18 & 92.73 & 94.92 & 0.6  & 0.45 & 2.1  & 0.38 & 0.2  & 0.26 \\
         &      & HMM  & 50.02 & 57.21 & 75.41 & 59.17 & 64.43 & 57.78 & 9.27 & 1.53 & 13.51 & 1.74 & 0.36 & 2.28 \\
         &      & Obs  & 62.29 & -    & 89.69 & 93.27 & 89.82 & 92.44 & -    & -    & -    & -    & -    & - \\\midrule
    20   & 200  & $D$-known & 77.52 & 89.05 & 94.33 & 95.1 & 92.36 & 94.71 & -    & -    & -    & -    & -    & - \\
         &      & Proposed & 77.1 & 87.74 & 94.1 & 94.94 & 92.19 & 94.62 & 2.8  & 2.56 & 21.17 & 1.26 & 0.51 & 1.62 \\
         &      & HMM  & 49.21 & 55.77 & 75.08 & 58.89 & 64.05 & 57.8 & 13.52 & 2.89 & 15.15 & 5.11 & 0.89 & 4.82 \\
         &      & Obs  & 52.61 & -    & 87.69 & 87.75 & 81.52 & 85.96 & -    & -    & -    & -    & -    & - \\
         & 500  & $D$-known & 78.05 & 89.35 & 94.55 & 95.21 & 92.53 & 94.94 & -    & -    & -    & -    & -    & - \\
         &      & Proposed & 77.91 & 88.8 & 94.48 & 95.17 & 92.47 & 94.92 & 3.32 & 1.94 & 14.97 & 0.43 & 0.22 & 0.62 \\
         &      & HMM  & 50.02 & 57.21 & 75.41 & 59.17 & 64.43 & 57.78 & 9.27 & 1.53 & 13.51 & 1.74 & 0.36 & 2.28 \\
         &      & Obs  & 56.24 & -    & 88.44 & 91.94 & 86.41 & 89.05 & -    & -    & -    & -    & -    & - \\\midrule
    500  & 200  & $D$-known & 78.14 & 89.18 & 94.43 & 95.48 & 92.28 & 94.54 & -    & -    & -    & -    & -    & - \\
         &      & Proposed & 77.42 & 87.81 & 94.15 & 95.33 & 92.18 & 94.37 & 2.73 & 2.5  & 21.79 & 1.19 & 0.52 & 1.67 \\
         &      & HMM  & 49.21 & 55.77 & 75.08 & 58.89 & 64.05 & 57.8 & 13.52 & 2.89 & 15.15 & 5.11 & 0.89 & 4.82 \\
         &      & Obs  & 53.83 & -    & 87.02 & 87.85 & 81.49 & 85.24 & -    & -    & -    & -    & -    & - \\
         & 500  & $D$-known & 79.06 & 89.55 & 94.55 & 95.61 & 92.41 & 94.78 & -    & -    & -    & -    & -    & - \\
         &      & Proposed & 78.98 & 89.37 & 94.62 & 95.61 & 92.51 & 94.8 & 2.22 & 1.51 & 10.33 & 0.4  & 0.21 & 0.47 \\
         &      & HMM  & 50.02 & 57.21 & 75.41 & 59.17 & 64.43 & 57.78 & 9.27 & 1.53 & 13.51 & 1.74 & 0.36 & 2.28 \\
         &      & Obs  & 56.74 & -    & 86.56 & 89.93 & 88.09 & 90.52 & -    & -    & -    & -    & -    & - \\
    \bottomrule
    \end{tabular}%
    }
    ~\\   {\scriptsize \raggedright \textbf{NOTES} 
        ``Accuracy \%'': overall classification accuracy in the test sets, where ``Posterior'' and ``Viterbi'' refer to the posterior probability prediction rule and the Viterbi's prediction rule, respectively; ``AUC$\times 10^2$'': the area under the ROC curve in the test set times 100 (using ``Posterior'' rule), where the numbers below refer to the classes that are regarded as ``correct'' in the respective one-versus-rest classifications; ``Bias$\times 10^3$'' and ``MSE$\times 10^3$'': the average bias and mean squared error for the free parameters in the HMM times 1000, where ``Tran'', ``Emis'', and ``Init'' refer to the transition intensity matrix, the emission matrix, and the initial state probability vector, respectively. {An entry with ``-'' indicates the quantity is not applicable for the corresponding method.}
    \par}
  \label{tab:cov1p4_tab1}\label{tab:cov1p20_tab1}%
\end{table}%

Results for $p=20$ and $500$ shown in \Cref{tab:cov1p20_tab1,tab:cov1p20_tab2} also demonstrate promising performance when the dimensionality of the observed variables becomes higher. In addition to the above several assessment measures, in \Cref{tab:cov1p20_tab2}, Column ``C'' represents the mean number of correctly identified signal variables (the ideal number is 4), ``IC'' represents the mean number of noise variables incorrectly identified as signals (the ideal number is 0){, and ``MCC'' represents an overall measure of the variable selection performance (the maximum Matthews correlation coefficient is 1). Regarding variable selection, our proposal has a slightly higher Matthews correlation coefficient than \textit{$D$-known}. An explanation may be that our method incorporates class probabilities in the multinomial regression, which provides additional information and a stochastic algorithm than the specific class labels in \textit{$D$-known}.}

\begin{table}[!htbp]
  \centering
  \caption{Logistic model parameter estimation and variable selection results with variable scenario (I)}
    \resizebox{\textwidth}{!}{
    \begin{tabular}{ccccccccccccccccc}
    \toprule 
    \multirow{2}[0]{*}{$p$} & \multirow{2}[0]{*}{$n$} & \multirow{2}[0]{*}{Method} & \multirow{2}[0]{*}{C} & \multirow{2}[0]{*}{IC} & \multirow{2}[0]{*}{MCC} & \multicolumn{4}{c}{Bias}  & \multicolumn{4}{c}{MSE}   & \multicolumn{3}{c}{MISD} \\\cmidrule(lr){7-10}\cmidrule(lr){11-14}\cmidrule(lr){15-17}
         &      &      &      &      &      & $\bbeta_{(1)}$ & $\bbeta_{(2)}$ & $\bbeta_{(3)}$ & $\bbeta_{(4)}$ & $\bbeta_{(1)}$ & $\bbeta_{(2)}$ & $\bbeta_{(3)}$ & $\bbeta_{(4)}$ & $\alpha_1$ & $\alpha_2$ & $\alpha_3$ \\\midrule 
    20   & 200  & $D$-known & 4    & 0    & 1    & 0.36 & 0.49 & 0.28 & 0.36 & 0.17 & 0.27 & 0.09 & 0.17 & 0    & 0    & 0 \\
         &      & Proposed & 4    & 0    & 1    & 0.58 & 0.61 & 0.35 & 0.46 & 0.40 & 0.40 & 0.14 & 0.27 & 3.74 & 3.62 & 3.86 \\
         & 500  & $D$-known & 4    & 0    & 1    & 0.26 & 0.39 & 0.22 & 0.29 & 0.09 & 0.17 & 0.06 & 0.10 & 0    & 0    & 0 \\
         &      & Proposed & 4    & 0    & 1    & 0.41 & 0.49 & 0.28 & 0.36 & 0.20 & 0.25 & 0.09 & 0.16 & 0.67 & 0.58 & 0.7 \\\midrule 
    500  & 200  & $D$-known & 4    & 23.97 & 0.60  & 0.12 & 0.11 & 0.09 & 0.11 & 0.06 & 0.09 & 0.04 & 0.06 & 0    & 0    & 0 \\
         &      & Proposed & 4    & 0.27 & 0.99 & 0.59 & 0.54 & 0.34 & 0.43 & 0.43 & 0.33 & 0.13 & 0.24 & 5.08 & 4.83 & 5.52 \\
         & 500  & $D$-known & 4    & 32.05 & 0.53 & 0.08 & 0.09 & 0.06 & 0.08 & 0.02 & 0.04 & 0.02 & 0.03 & 0    & 0    & 0 \\
         &      & Proposed & 4    & 4.43 & 0.90  & 0.30 & 0.29 & 0.19 & 0.25 & 0.12 & 0.12 & 0.05 & 0.09 & 0.82 & 0.91 & 0.82 \\
    \bottomrule 
    \end{tabular}%
    }
    ~\\   {\scriptsize \raggedright \textbf{NOTES}  
        ``Bias'' and ``MSE'': the average bias and mean squared error for the logistic parameters; ``MISD'': the average mean integrated squared difference for the time-varying intercepts where \textit{$D$-known} is used as the benchmark; ``C'': mean number of correctly identified signal variables (the ideal number is 4); ``IC'': mean number of noise variables incorrectly identified as signals (the ideal number is 0); ``MCC'': Matthews correlation coefficient of variable selection.%
    \par}
  \label{tab:cov1p4_tab2}\label{tab:cov1p20_tab2}%
\end{table}%

\subsection{Sensitivity Analysis}\label{sec:sensitivity}
We present four sets of sensitivity analyses: In (A), we investigate the case where none of the disease markers $\bX$ is informative. In (B), we assess the performance of the proposed method when the emission probability of $Z$ given $D$ is a function of $\bX$. In (C), we change the data-generating scheme for the observed variables so that the logistic model no longer holds. In (D), we assess the validity of the prediction rules when the underlying prevalence of diseases is different among the training and the test sets. Details of the data generation are deferred to Appendix E. 
In general, the results are similar to the primary simulation studies, which show that our proposal is robust to violations of the data-generating schemes. 

Tables S4 and S5 in the Supplementary Material summarize the prediction results of the sensitivity analysis. 
For (A), as expected, \textit{$D$-known} and \textit{Proposed} perform similarly, as noise variables do not contain any information for the prediction{; \textit{HMM} provides higher accuracies since it still uses the surrogate labels in the testing set}. Results for (B) suggest that when the emission distributions of $Z$ are misspecified, our proposal tends to approximate them as time-averaged values. Despite this misspecification, the other progression parameters appear to remain well estimated, and the prediction accuracy is largely unaffected. Setting (C) indicates that even if the underlying logistic model is not satisfied, the discriminative model can still serve as a reliable prediction rule. %
{Results of Setting (D) show that the accuracy of \textit{Proposed} is largely unaffected under $p=4$, but using Viterbi's algorithm leads to a larger decrease in the accuracy under the more challenging case of $p=20$. We point out the reason and discuss the remedy in Appendix E.} 
Furthermore, Tables S6 and S7 show that the proposed method can estimate the logistic model parameters well even when the underlying model assumptions are violated.%

\section{Application to NACC Study}\label{sec:realdata}

Our analysis aims to construct a time-sensitive diagnostic rule to distinguish LBD-only or AD\&LBD from AD-only at an early stage for future subjects during their lifespans. In the NACC study, neuropathological diagnoses are {available in many participants after death, while} %
antemortem assessments, including subjects' motor, behavioral, and cognitive functioning, as well as surrogate clinical diagnoses, are available {for multiple visits preceding death.} %
The history of all these measures will be used in the prediction.  We follow \cite{mcdonnell2021dynamic} to process disease markers in our proposed model. The time-invariant variables include %
{the number of APOE e4 alleles} and demographic variables%
, and time-varying variables include measures in the motor, cognitive, and psychiatric domains. {We did not have sufficient information in this dataset to include glucocerebrosidase %
or other non-APOE genes, plasma or cerebrospinal fluid biomarkers (A$\beta$42, A$\beta$40, phosphotau, synuclein seed tests), or magnetic resonance imaging, beta-amyloid positron emission tomography, or dopamine transporter imaging biomarkers.} {Details of the variables and pre-processing are presented in Appendix F}. %
The final dataset includes 1,875 individuals {who underwent autopsy}, among whom 1,247 had {gold-standard} neuropathological diagnosis as AD-only, 155 as LBD-only, and 473 as AD\&LBD. The average age at the first visit was 74.5%
, and the average follow-up %
was 5.5 %
years. The average number of visits per individual was 5.7%
, contributing to a total of 10,754 observations.

To evaluate the prediction performance, we randomly partition the data into a training set and a testing set and %
include fewer pathologically AD-only patients in the testing data so that it will not dominate the performance. The disease prevalences in the test set are %
45.2\%, 13.1\%, and 41.7\% %
for AD-only, LBD-only, and AD\&LBD, respectively. To evaluate the performance of classifiers, we only use the observations at the last time point %
as the gold-standard neuropathological diagnosis is only available at that time. We  
focus on classifying individuals to the most probable disease state (i.e., AD-only, LBD-only, or AD\&LBD). We apply our method with the adaptive lasso penalty and compare it with other competitors similar to those used in the simulation studies. %
{Since Viterbi-based prediction uses all the history of ${\bX}(t)$ and generally outperforms posterior-based prediction, as shown in the simulation studies, we consider Viterbi-based prediction for both the \textit{HMM} and our \textit{Proposed} method. {In fact, clinical diagnoses are made at longitudinal follow-up visits, and consensus diagnosis based on the entire longitudinal profile may improve accuracy as observations accrue. Viterbi algorithm offers a viable solution to leverage this effect.} %

\begin{table}[!htpb]
    \centering
    \caption{Prediction accuracy and one versus all classification performance at the last observation time. 
    The observations at the last time points are classified to the most probable disease state (i.e., AD-only, LBD-only, or AD\&LBD). 
    The proposed method has the highest accuracy in predicting ADRDs, and performs better in distinguishing LBD-only versus AD-only. \textit{Obs} tends to classify all patients to AD-only.}
    \label{tab:NACCaccuracy}
    \resizebox{\textwidth}{!}{
    \begin{tabular}{cccccccccc}
    \toprule 
         & \multicolumn{3}{c}{Proposed} & \multicolumn{3}{c}{Obs} & \multicolumn{3}{c}{HMM} \\
    Overall Accuracy & \multicolumn{3}{c}{\textbf{48.3}} & \multicolumn{3}{c}{47.3} & \multicolumn{3}{c}{42.9} \\
    \textit{``One vs all'' performance} & AD-only & LBD-only & AD\&LBD & AD-only & LBD-only & AD\&LBD & AD-only & LBD-only & AD\&LBD \\ 
    Balanced Accuracy & \textbf{0.57} & \textbf{0.76} & \textbf{0.53} & 0.55 & 0.53 & 0.50 & 0.52 & 0.65 & 0.50 \\
    Sensitivity & 0.50 & \textbf{0.64} & 0.42 & \textbf{0.98} & 0.08 & 0.05 & 0.39 & 0.41 & \textbf{0.47} \\
    Specificity & 0.64 & 0.87 & 0.65 & 0.13 & \textbf{0.98} & \textbf{0.94} & \textbf{0.65} & 0.89 & 0.52 \\
    PPV & \textbf{0.53} & \textbf{0.43} & \textbf{0.46} & 0.48 & 0.40 & 0.39 & 0.48 & 0.35 & 0.41 \\
    NPV & 0.60 & \textbf{0.94} & \textbf{0.61} & \textbf{0.87} & 0.88 & 0.58 & 0.57 & 0.91 & 0.58 \\
    \bottomrule 
    \end{tabular}%
    }
    ~\\   {\scriptsize \raggedright \textbf{NOTES}  
        ``Balanced Accuracy'': average of sensitivity and specificity; ``PPV'' and ``NPV'': positive and negative predictive values, respectively.
    \par}
\end{table}

The prediction accuracy and three ``one versus all'' classifications performance are reported in \Cref{tab:NACCaccuracy}. The 
proposed method achieves the highest accuracy in the overall classification and well-balanced ``one versus all'' accuracy in the three binary classifications compared to the other methods. 
Furthermore, the proposed method has a comparable sensitivity to \textit{HMM} for distinguishing AD\&LBD from AD-only, but a much higher sensitivity for distinguishing LBD-only from AD-only. In contrast, the pointwise-based method, i.e., \textit{Obs}, often misclassifies LBD-only and AD\&LBD patients as AD-only. Moreover, the proposed method yields the highest positive and negative predictive values in almost all comparisons. Overall, the proposed method strikes a good balance among different metrics, indicating better identification of true disease states.

We present variables selected as informative of distinguishing {LBD-only and AD\&LBD} from AD-only and their coefficients in Table S9. {The proposed method effectively balances classification accuracy and variable selection as it provides a more parsimonious model.} %
Our results indicate that the following variables are most useful in discriminating LBD-only from AD-only: 
less impairment in memory, more severe hallucination, and fewer copies of APOE e4 alleles. Our results are consistent with current clinical literature, which supports hallucination as a key indicator of LBD \citep{yoshizawa2013early,mckeith2020research} and memory impairment is often observed in AD patients but less so in LBD patients \citep{salmon1996neuropsychological}. Other important indicators of LBD include less impairment in orientation{, higher impairment in community affairs,} and needing more personal care. Most variables manifest a weaker strength in distinguishing AD\&LBD from AD due to the shared pathology between the two groups. However, the AD\&LBD group is more likely to be males and present more severe hallucinations than the AD-only group. Using clinical diagnosis would indicate less memory impairment in AD\&LBD group compared to AD-only {(the estimate is -0.79)}, which is inconsistent with empirical evidence{, i.e., memory impairment is similar for neuropathologically AD\&LBD and AD-only at the last clinical visit (1.32 vs 1.33, respectively)}. 

{\mbox{\Cref{fig:cov_trajectory}} shows the mean trajectory of several markers over time for each disease state predicted from the Viterbi algorithm using the proposed method. The results show a clear separation between LBD-only and the other classes across time for multiple measures in the Clinical Dementia Rating (e.g., memory, orientation) and functional activities. In contrast, it is more difficult to distinguish AD-only and AD\&LBD (apathy severity has some discriminative power when approaching death).}

\begin{figure}[!htbp]
    \centering
    \includegraphics[width=\linewidth]{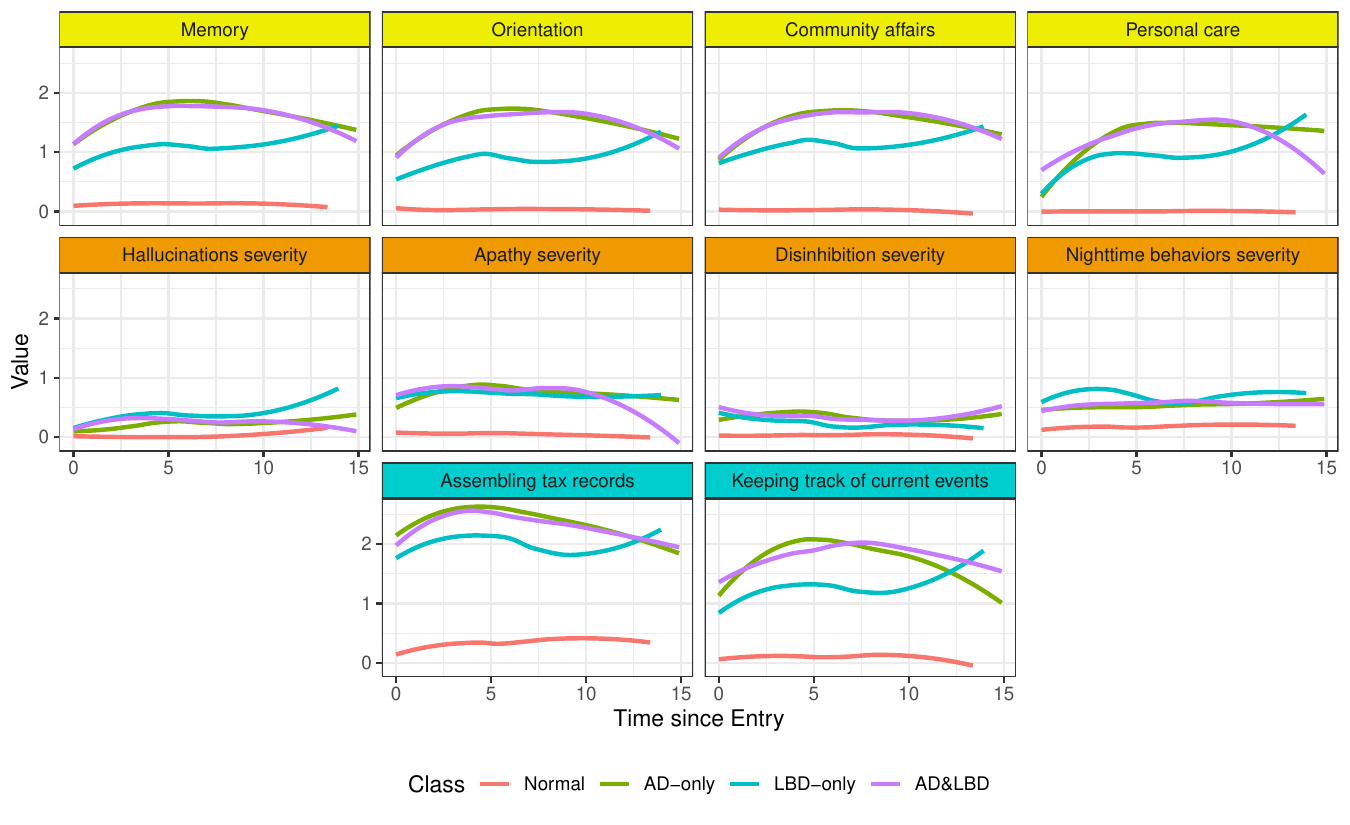}
    \caption{Average disease marker trajectory (smoothed with locally estimated scatterplot smoothing) over time for each group predicted by the proposed method with the Viterbi algorithm. Measures from the Clinical Dementia Rating (CDR® Plus FTLD), Neuropsychiatric Inventory Questionnaire, and Functional Assessment Scale are in yellow, orange, and cyan, respectively.
    The CDR and FAS separate LBD-only and the other classes across time well, while it is more difficult to distinguish AD-only and AD\&LBD.
    }
    \label{fig:cov_trajectory}
\end{figure}

\section{Discussion}\label{sec:discussion}
The article establishes the methodological and theoretical foundations for a new class of generative-discriminative modeling framework aiming at improving the prediction and forecast of progression by leveraging both objective information ($\bX$) and surrogate labels ($Z$). The proposed method addresses the limitations of relying on highly variable clinical diagnoses by focusing on more reliable objective measures for future predictions. The framework is designed to separate these two sources of information, allowing for a prediction rule that uses only $\bX$ for predicting future disease progression but contains information from already observed clinical diagnoses. Furthermore, generative modeling allows us to capture disease mechanisms and simulate new patient's disease outcome to assist design of future clinical trials.

The proposed method can be further extended in several directions. The generative component can be extended to accommodate more complex disease progression patterns. For example, heterogeneity in transition intensity can be introduced by incorporating individual characteristics that are not direct manifestations of the underlying states. Using a higher-order Markov chain on true disease states could capture more complex temporal dependencies. Allowing increasing the number of latent states, such as incorporating disease sub-types, would enable a finer representation of the underlying process. It is also of interest to investigate settings where the number of classes is unknown and needs to be estimated or where the underlying states are in a continuous space. 

The discriminative model can also be extended to capture more dynamic effects of the disease states on disease markers, as indicated in \Cref{sec:EM}. While we consider the fixed $p$ regime in this article, the methodology and theory can be generalized to ultra-high dimensions where $p_n$ grows with $n$. Furthermore, non-Euclidean emission of $X$, like text or images, can also be incorporated in the discriminative component via embedding into a lower-dimensional Euclidean space using deep learning techniques. These directions are left for future exploration.

\section*{Competing interests}
No competing interest is declared.

\section*{Acknowledgement}
This research is supported by the U.S. National Institutes of Health (NS073671, GM124104, MH123487, U01NS100600, U24AG072122, UL1TR001873) and LBDA RCOE CU.

\section*{Supplementary material}
\label{SM}
The Supplementary Material includes technical lemmas, proof of the lemmas and theorems, and additional simulations and real data analysis results.

\bibliographystyle{apalike}

\end{document}